\documentstyle[psfig,preprint,aps]{revtex}

\begin{document}
\tighten
\pssilent
\preprint{\vbox{
\hbox{SISSA Ref. 18/96/A}
\hbox{CERN-TH/95-351}
\hbox{IC/96/17}
\hbox{\bf hep-ph/9601376}}}

\title{NONRESTORATION OF SPONTANEOUSLY BROKEN P,  
CP AND PQ AT HIGH TEMPERATURE}

\author{{\bf Gia  Dvali}\thanks{ e-mail adresses:
 dvali@surya11.cern.ch, melfo@stardust.sissa.it, goran@ictp.trieste.it }}

\address{CERN, CH-1211, Geneva 23, Switzerland}
\author{{\bf Alejandra Melfo} }
\address{SISSA,  34014 Trieste, Italy,
 {\rm and}  Centro de Astrof\'{\i}sica Te\'orica, Universidad de Los
Andes M\'erida 5101-A, Venezuela}
\author{{\bf Goran Senjanovi\'c}}
\address{International Center for Theoretical Physics
34100 Trieste, Italy }
\date{30/01/96}
\maketitle

\begin{abstract}

The possibility of P and  CP violation at high temperature in models where
 these symmetries  are spontaneously broken is investigated. It is found 
that in  minimal models that include singlet fields, high T nonrestoration 
is possible for a wide range of parameters of the theory,
in particular in models of CP violation with a CP-odd Higgs field.
The same holds true for   the invisible axion version of the Peccei-Quinn 
mechanism. This can provide both
 a way out for the domain wall problem in these theories and the
 CP violation required for baryogenesis. 
In the case of spontaneous P violation it turns out that high T 
nonrestoration requires going beyond the minimal model.
The results are shown to hold true when next-to-leading order effects are 
considered. 

\end{abstract}

\newpage

\section{Introduction}
\label{intr}

The phenomenon of spontaneous symmetry breaking has become a cornerstone of
 modern particle physics. To be able to establish a connection between 
particle physics and cosmology, it is essential to investigate the behavior
 of symmetry breaking in the early universe, {\em i.e.} at high temperature.
In spite of common sense prejudice, it is by now known that more heat does 
not necessarily imply more symmetry \cite{w74,ms79}. 
Rather, the question of symmetry 
restoration is quite a complex phenomenon and depends on the dynamics of
 the theory considered. 

Examples have been found with symmetries remaining
 broken at arbitrarily high temperature, or even exact symmetries becoming
 broken as the system gets heated up \cite{ms79,lp80,kst90,dw90}. 
However, some of these examples were
artificially created just in order to demonstrate the phenomenon. In
 our opinion, symmetry nonrestoration  becomes relevant  only
when resulting from minimal and realistic models. 
This is precisely what we wish to address in this paper.
 For the sake of focus, we concentrate on the issues of P and CP
 violation (both weak and strong). The choice of parity and time reversal
 is in our opinion natural, 
these being fundamental symmetries of nature. Furthermore, the 
spontaneous breaking of these symmetries may offer a simple way out of 
the strong CP problem \cite{h78}.

There are at least two important reasons to have CP broken at
 high temperature. Baryogenesis requires  CP violation, and if one is to
 adhere to the appealing idea of CP symmetry being broken  spontaneously, its
 nonrestoration becomes a must.  On the other hand, the spontaneous 
breakdown of a discrete symmetry leads to a domain wall problem, 
following the phase transition that takes place if the symmetry is 
restored 
at high T \cite{zko74,k76}. Avoiding this phase transition may be 
sufficient to
 solve 
the problem, since the thermal production of large domain walls is 
naturally suppressed for a wide range of the parameters of the theory 
\cite{ds95}. In section \ref{weakcp}, we study CP behavior at high temperature
in some  $SU(2)\times U(1)$ theories with Higgs doublets and singlets 
only. It turns out that in minimal such models with doublets only 
CP is always restored, whereas it can naturally remain broken if there 
is at least one singlet on top of the usual Higgs doublet.

Section \ref{p} is devoted to P violation and there we find that 
nonrestoration of P at high T seems to be in conflict with perturbation theory.
Again, the existence of P odd singlets, welcome for the implementation of
 the minimal see-saw mechanism, works in favor of nonrestoration of P 
just as in the case of CP. 

There is yet another class of theories  plagued by the domain wall 
problem, that is, those based on the Peccei-Quinn solution 
\cite{pq77} to the strong CP problem. Once again, symmetry nonrestoration
 can solve the problem \cite{ds95}.
 In section \ref{strongcp} we demonstrate in detail how this is achieved.

It has been  pointed out that next-to-leading order corrections to the high
 temperature effective potential may play an important role on the question
of nonrestoration, even to the extent of invalidating it  in the case of
 local gauge symmetries \cite{bl95a,bl95b}. However, a  recent study \cite{r95}
involving a Wilson renormalization group approach which simulates 
nonperturbative effects, seems to encourage the validity of the 
conventional one loop results. Since the issue is not completely settled,
to be on the safe side we show in section \ref{ntl} how inclusion of 
next-to-leading order terms does not affect any of our conclusions.

Focusing on CP forced us to ignore some rather important applications of the
idea of symmetry nonrestoration, in particular a possible solution to the
 monopole problem in grand unified theories \cite{sss85,nos}. We leave this 
and related issues for the future.

\section{Spontaneous CP Violation and High T}
\label{weakcp}

As with any discrete symmetry, we would like to be able to keep CP broken at 
high temperature in order to avoid the formation of the dangerous domain walls.
In the case of CP, there is yet an additional reason not to restore it in 
the early universe, at least not until the time of baryogenesis. Simply, CP
 must be broken in order for matter to be created \cite{s67}. This was 
actually the original motivation of the first application in  particle 
physics of the phenomenon of nonrestoration of symmetries at high temperature
\cite{ms79}. The model presented in \cite{ms79} however does not satisfy 
the minimality condition introduced above, since there the Higgs sector  
is extended  to three doublets only in order  to have high T symmetry 
nonrestoration.

\subsection{$\; \not\!\!\!\!{\rm CP}$ with two doublets}
\label{tdl}

The simplest and original example of a theory with spontaneous CP violation 
was presented by T.D. Lee \cite{l73}. His model is an  
extension of the Standard Model with two complex Higgs doublets, with

\begin{equation}
{\cal L}_{H} = \sum_{i=1}^2{{1 \over 2}(D_\mu \Phi_i)^\dagger (D^\mu \Phi_i)} 
- V(\Phi_1, \Phi_2)
\label{tdlag}
\end{equation}

where

\begin{eqnarray}
 V(\Phi_i, \Phi_2) &=& \sum_{i=1}^2{\left(- {m_i^2 \over 2} 
\Phi_i^\dagger\Phi_i
 + {\lambda_i \over 4} (\Phi_i^\dagger\Phi_i)^2 \right)} - {\alpha \over 4} 
\Phi_1^\dagger\Phi_1\,\Phi_2^\dagger\Phi_2 \nonumber \\
&-& {\beta \over 4} \Phi_1^\dagger\Phi_2\,\Phi_2^\dagger\Phi_1 + {1 \over 8} 
\left[ 
\Phi_1^\dagger \Phi_2 \left(a \Phi_1^\dagger\Phi_2 + b\Phi_1^\dagger\Phi_1+ c 
\Phi_2^\dagger\Phi_2 \right) + \, h.c. \right] 
\label{tdlpot}
\end{eqnarray}

Choosing the parameter $\beta >0$, one can prove that the minimum of 
the potential is achieved when the fields acquire vevs

\begin{equation}
\Phi_1 =\left(\begin{array}{c} 0\\ v_1 \end{array} \right) \; \; \; \; ; \;\;\;
\Phi_2 =\left(\begin{array}{c} 0\\ v_2 \end{array} \right) e^{i\theta}
\label{tdlvev}
\end{equation}

The terms in brackets  in the potential will force the 
CP-violating phase $\theta$ to be non-zero. This can be readily seen by 
writing (\ref{tdlpot}) at the minimum 
(\ref{tdlvev}), and wisely rearranging terms:

\begin{equation}
V(\langle \Phi_1 \rangle,\langle \Phi_2 \rangle) = \sum_{i=1}^2{\left(- {m_i^2
 \over 2} v_i^2 + {p_i \over 4} v_i^4 \right)} + {\rho \over 4} v_i^2 v_2^2 +
 {a \over 2} 
v_i^2 v_2^2 
[\cos\theta - \delta]^2 
\label{tdlmin}
\end{equation}

where

\begin{eqnarray}
p_1 = \lambda_1 - {b^2 \over 8 a} \;\;\;\; &;&  \;\;\;\;
p_2 = \lambda_2 - {c^2 \over 8 a} \nonumber \\
\rho = \alpha +\beta + a + {cb \over 4 a} \;\;\;\; &;&  \;\;\;\;
\delta = {-(b v_1^2 + c v_2^2) \over 4 a v_1 v_2}
\label{trick}
\end{eqnarray}

Obviously, for $a>0$ the minimum will be at $\cos\theta =\delta$, 
and CP is broken spontaneously. 

We are interested in the possibility that CP remains
 broken at arbitrarily high temperature. For this to happen in T.D. Lee's
 model, we need  not only to have the vev's of {\em both} $\Phi_1$ and
 $\Phi_2$ nonzero at high T, but also to keep the CP-violating phase 
 from vanishing.

To get an idea of how both vev's may be kept different from zero, 
consider a simple model with two real scalar fields ($\phi_1, \phi_2$), 
and a potential with a $Z_2$ symmetry 
$\phi_1\rightarrow - \phi_1$, $\phi_2\rightarrow - \phi_2$

\begin{equation}
V(\phi_1, \phi_2) = \sum_{i=1}^2\left(- {m_i^2 \over 2} \phi_i^2 +
 {\lambda_i \over 4} \phi_i^4 \right) - {\alpha \over 2} \phi_1^2 \phi_2^2 
+ \beta_1\, \phi_1^3 \phi_2 + \beta_2\, \phi_2^3 \phi_1
\label{toy}
\end{equation} 

One can always choose $\alpha >0$, $\beta_1, \beta_2 >0$, and require 

\begin{equation}
\lambda_1 \lambda_2 > \alpha^2
\label{bound}
\end{equation}

so that the potential is bounded from below. The potential has extrema at 
$\langle \phi_1 \rangle = v_1$, 
$\langle \phi_2 \rangle = v_2$ satisfying

\begin{mathletters}
\label{toymin}
\begin{equation}
[-m_1^2 + \lambda_1 v_1^2 - \alpha v_2^2 + 3\beta_1 v_1 v_2] v_1 + 
\beta_2 v_2^3 =0
\end{equation}
\begin{equation}
[-m_2^2 + \lambda_2 v_2^2 - \alpha v_1^2 + 3\beta_2 v_2 v_1] v_2 +
 \beta_1 v_1^3 =0
\end{equation}
\end{mathletters}

With negative mass terms, both vevs are nonzero. Admittedly, this model does
 not belong to the class of minimal models as defined in this paper, since
 one can break the $Z_2$ symmetry with just one vev; however, we include it
 in order to illustrate the role of the linear terms in symmetry
 nonrestoration. 

At high temperature, the 
effective potential acquires the additional terms \cite{w74,kl72,dj74}

\begin{equation}
\Delta V = {T^2\over 24} \left[(3 \lambda_1 - \alpha)\phi_1^2 + (3 \lambda_2 - 
\alpha)\phi_2^2 + 6 (\beta_1 + \beta_2) \phi_1 \phi_2 )\right]
\end{equation}

By asking, e.g. $\alpha > 3 \lambda_1$, one can keep one of the mass terms 
negative at any temperature, while (\ref{bound}) forces the other to 
be positive. However,  the cubic terms in (\ref{toymin}) guarantee that 
only one negative mass term suffices to have {\em both} vevs non zero 
at high T. In other words, the field with the negative
 mass term acquires a vev and ``forces'' the other to get one also, via the
 linear terms in the potential. The reader must have noticed that we can
 redefine the fields at high T so that just one of them has a nonvanishing
 vev. However, she should keep in mind that the same holds true at $T=0$;
 the point is that the symmetry breaking patterns at high and low T 
are equal. 

One can hope that the potential in T.D. Lee's model, being of a similar form
 as (\ref{toy}), will exhibit a similar behavior, allowing both vev's to remain
 non zero at high temperature. Unfortunately, it is readily found out that 
it does so at the expense of 
having the phase $\theta$ going to zero, thus restoring CP, as we now show.

The high temperature corrections to the effective potential for a model 
with N Higgs doublets can be found by generalizing Weinberg's 
formula \cite{w74} for complex doublets. 
Write the most general potential for N complex doublets as 
\footnote{Obviously we do not worry about the potential being hermitian. 
Needless to say, the reader should take care of this in choosing her 
potential, and then safely proceed to use our formula for $\Delta V(T)$}

\begin{equation}
V = -\sum_{i=1}^N{ m_i^2 \Phi_i^\dagger \Phi_i} + \sum_{i,j,k,l=1}^N{
\lambda_{ijkl}
\Phi_i^\dagger \Phi_j \Phi_k^\dagger \Phi_l}
\end{equation}

Then the high T correction is

\begin{equation}
\Delta V(T) = \sum_{i,j,k=1}^N{ {T^2 \over 6} \left(2 \lambda_{ijkk}
 + \lambda_{kijk} \right)\Phi_i^\dagger \Phi_j}
\label{Tdmass}
\end{equation}

For the two doublet model (\ref{tdlpot}), this gives

\begin{equation}
\Delta V(T) = {T^2 \over 6} \left[ (6 \lambda_1 - 2 \alpha - \beta)
 \Phi_1^\dagger \Phi_1 + (6 \lambda_2 - 2 \alpha - \beta)
 \Phi_2^\dagger \Phi_2  
+ {3 \over 2} (b+c) (\Phi_1^\dagger \Phi_2  + h.c )\right]
\end{equation}

The potential at high T can then be cast in the same form (\ref{tdlmin}), 
where now the masses $m_i^2$ are replaced by $m_i^2(T)$

\begin{eqnarray}
m_1^2(T) &=& - m_1^2 + 2 T^2 \left(\lambda_1 - {\alpha\over 3} -
{\beta \over 6} - {b (b+c) \over 16 a}\right) \simeq 2 T^2 \nu_1^2
  \nonumber \\
m_2^2(T) &=& - m_2^2 + 2 T^2 \left(\lambda_2 - {\alpha\over 3} -
{\beta \over 6} - {c (b+c) \over 16 a}\right)  \simeq 2 T^2 \nu_2^2
\end{eqnarray}

for $T>>m$;  and $\delta$ becomes $\delta(T)$:

\begin{equation}
\delta(T) = -\left[{b v_1^2 + c v_2^2 + T^2 (b+c) \over 4 a v_1 v_2 }\right]
\end{equation}

Again, as in the simpler  model, one can have one and only one mass
 negative at high T, due to the condition analogous to  (\ref{bound}), i.e.

\begin{equation}
p_1 p_2 >{ \rho^2 \over 4}
\label{Tbound}
\end{equation}
since now
\begin{equation}
\nu_1^2 = p_1 - \sigma  \; ;\hspace{0.5cm}   \nu_2^2 = p_2 - \sigma 
 \;; \;\; \; 
{\rm with} \hspace{0.5cm}
\sigma = {\rho \over 2} - {\alpha \over 6} - {\beta \over 3} - 
{a \over 2} < { \rho\over 2}
\end{equation}

Requiring $\nu_1^2, \nu_2^2 < 0 $ will give $\,p_1 p_2 < \sigma^2
 < \rho^2 /4\,$, 
which contradicts (\ref{Tbound}). 

Considering only the $\theta$-dependent part, we see as before that there 
is a minimum for $\theta = \delta(T)$. However, it is not difficult 
to see that with only one mass term negative, both vevs cannot be nonzero 
at high T, due to the fact that the mass terms now depend on the coupling 
constants. Taking $\nu_2^2 < 0$, the requirement that $v_1 $ be real gives

\begin{equation}
|\nu_2^2| {\rho \over 2} > \nu_1^2 p_2
\label{realv1}
\end{equation} 

together with (\ref{Tbound}), this is also enough to ensure that $v_2$ is 
real. Substituting for $\nu_1^2$ and $\nu_2^2$ one gets

\begin{equation}
{\rho \over 2} \left({\rho \over 2} -p_2 \right)>{\rho \over 2}
 \left(\sigma -p_2 \right) > \left(p_1 -\sigma \right) p_2
> \left(p_1 -  {\rho \over 2} \right) p_2
\end{equation} 

Which again implies $\,p_1 p_2 < \rho^2 /4\,$, contradicting  (\ref{Tbound}).

We conclude then that the only way to have both fields with a nonvanishing
 vev at high temperature is to set the phase $\theta$ to zero. In other words, 
the field with a negative mass term can ``force'' the other to acquire a vev, 
but it drags it in the same direction in $U(1)$ space.

Notice that in \cite{ms79} the fact that both vevs can be nonzero was
 overlooked, but it was still concluded correctly that with two doublets
 only, CP would become a good symmetry at high T.

\subsection{$\; \not\!\!\!\!{\rm CP}$ and natural flavor conservation}
\label{nfc}

A common feature of models with two Higgs doublets as the one in the previous 
section is that they allow for flavor-violating interactions in neutral 
current phenomena. 
As shown in \cite{w76,dm77,b80}, the minimal model for
 spontaneous CP violation involving  doublets only  that conserves 
flavor,  requires three of them. 

To see why, consider a Lagrangian with two complex Higgs as in(\ref{tdlag}), 
(\ref{tdlpot}), and an extra symmetry $D_1$
\begin{equation}
\Phi_1 \longrightarrow -\Phi_1 \hspace{1cm} u_{iR}\longrightarrow - u_{iR}  
\end{equation}

(where $u_{a \,R}$ are up quarks and hereafter $a, b,.. $ are  flavor indices).
 The Yukawa  interactions are written now

\begin{equation}
{\cal L}_Y =   (\bar u \bar d)_L^a h^1_{ab} \Phi_1 d^b_{R}+
(\bar u \bar d)_L^a h^2_{ab} (i\tau_2)\Phi_2^* u^b_{R} 
 \end{equation}

so that flavor violation through neutral Higgs exchange is avoided. However,
 now the symmetry prohibits the 
 terms  of the type $\Phi_1^\dagger \Phi_1 \Phi_1^\dagger \Phi_2$ 
in the Higgs potential, and therefore at the minimum  we have 
the phase $\theta = 0$ or $\pi/2$, both leading to CP conservation.

The way out is to have three doublets, and an additional  symmetry $D_2$
that prevents it from coupling to the quarks: 
$\Phi_3 \rightarrow -\Phi_3$, with other fields unchanged. The most general
potential invariant under $SU(2)\times U(1)\times D_1\times D_2$ is 

\begin{eqnarray}
V &= &  \sum_{i=1}^3{\left[- m_i^2 \Phi_i^\dagger \Phi_i +  \lambda_{i} 
 (\Phi_i^\dagger \Phi_i)^2 \right] }  \nonumber\\
& + &  \sum_{i<j}\left[ - \alpha_{ij}(\Phi_i^\dagger \Phi_i)
 (\Phi_j^\dagger \Phi_j)
- \beta_{ij}(\Phi_i^\dagger \Phi_j) (\Phi_j^\dagger \Phi_i) +
 \gamma_{ij}(\Phi_i^\dagger \Phi_j \Phi_i^\dagger
 \Phi_j + h.c.)\right]
\label{nfcpot}
\end{eqnarray}

It can be shown \cite{w76,dm77,b80} that choosing $\beta_{ij}, \gamma_{ij} >0,
 $ the above potential has a minimum at

\begin{equation}
\Phi_i = {1 \over \sqrt{2}} \left( \begin{array}{c}0 \\v_i e^{i \theta_i} 
\end{array}\right)
\end{equation}

where only two of the $\theta_i$ (say, $\theta_1$ and $\theta_3$) are relevant.
Extremization with respect to $\theta$ yields \cite{dm77}

\begin{mathletters}
\label{nfcmin}
\begin{equation}
\gamma_{12} v_2^2 \sin 2\theta_1 + \gamma_{13} v_3^2 \sin 2 (\theta_1 - 
\theta_3) =0
\end{equation}
\begin{equation}
\gamma_{13} v_1^2 \sin 2(\theta_1 - \theta_3) + \gamma_{23} v_2^2 \sin 2
 \theta_3 =0
\end{equation}
\label{nfcphase}
\end{mathletters}

Notice that to have CP violation, we need all three $v_i$ and both
 $\theta_1,\; \theta_3$ to be nonzero.

 It can  be 
shown \cite{b80} that the CP violating solution 
of (\ref{nfcmin}) is indeed a minimum. 
When the phases take this value, the remaining potential is

\begin{equation}
V(v_i) = \sum_{i=1}^3\left(- {m_i\over 2} v_i^2 + {p_i \over 4} v_i^4
\right) - \sum_{i<j}{(\alpha_{ij} + \beta_{ij}) \over 4} v_i^2 v_j^2
\label{nfcrem}
\end{equation}

where

\begin{equation}
p_1 = \lambda_1 - {\gamma_{12} \gamma_{13} \over \gamma_{23}}
\end{equation}

and analogous expressions for $p_2,\  p_3$.

Once again, we are interested in whether the CP symmetry can remain broken 
at high 
temperatures. It is straightforward using (\ref{Tdmass}) to calculate the 
masses 
at high temperature

\begin{equation}
m_i^2(T) = -m_i^2  + {T^2 \over 6}\left[6 p_i 
- \sum_{j\neq i}\left(2 \alpha_{ij} 
+ \beta_{ij}\right) \right] \simeq {T^2 \over 3} \nu_i^2
\label{nfcmass}
\end{equation}

Due to the high degree of symmetry of the potential, temperature 
contributions are independent of the phases, so equations  (\ref{nfcmin})
 are the same. 

For the potential to be bounded from below, a set of constraints analogous to
 (\ref{bound}) has to be imposed on the couplings, namely

\begin{mathletters}
\label{nfcbound}
\begin{equation} 
p_i>0 \hspace{1cm} p_i p_j > a_{ij}   \hspace{1cm} {\rm for\; each} \; i<j
\label{sdet}
\end{equation}
\begin{equation}
p_1 p_2 p_3 - p_1 a_{23}^2
 - p_2 a_{13}^2 - p_3 a_{12}^2 - 2 a_{12} a_{13} a_{23} >0
\label{det}
\end{equation}
\end{mathletters}

with $a_{ij} \equiv \alpha_{ij} + \beta_{ij}$, and we choose $\alpha_{ij} >0$,
 so $a_{ij}>0$. 

It is easy  to prove that (\ref{sdet}) prevents us from taking all 
three of the mass terms negative at high T, as we could have expected. 
Necessary conditions would be

\begin{equation}
\sum_{j\neq i} a_{ij} > 3 p_i
\end{equation}

Multiplying these equations by pairs and adding them results in  a 
contradiction with eq. (\ref{sdet}).
But it turns out that with only two negative mass terms, 
all three vevs cannot be nonzero at arbitrarily high temperature. 
Take for example $\nu_1^2 >0$, $\nu_2^2, \nu_3^2 <0$. We need $v_1$ to be real,
 that is, minimizing (\ref{nfcrem})

\begin{equation}
v_1^2  = \left({T^2 \over 3}\right) \frac{-\nu_1^2(p_2 p_3 - a_{23}^2) + 
\nu_2^2(p_3 a_{12} + a_{23} a_{13}) 
+ \nu_3^2(p_2 a_{13} + a_{23} a_{12})}
{p_1 p_2 p_3 - p_1 a_{23}^2
 - p_2 a_{13}^2 - p_3 a_{12}^2 - 2 a_{12} a_{13} a_{23}} >0
\label{nfcvev}
\end{equation}

We have already required the denominator to be positive. For the numerator to 
be positive also, necessary (though not sufficient) conditions are

\begin{equation}
\bar{\nu_2}^2(p_3 a_{12} + a_{23} a_{13}) +\bar{ \nu_3}^2(p_2 a_{13} + 
a_{23} a_{12}) > -\bar{\nu_1}^2(p_2 p_3 - a_{23}^2) 
\label{nfcn1}
\end{equation}

where
\begin{eqnarray}
\bar{\nu_1}^2 &=& 3 p_1 - a_{12} - a_{13}  <\nu_1^2 \nonumber \\
\bar{\nu_2}^2 &=&  a_{12} + a_{23} - 3 p_2 > \nu_2^2 \nonumber \\
\bar{\nu_3}^2 &=&  a_{13} + a_{23} - 3 p_3 >\nu_3^2  
\label{nfcn2}
\end{eqnarray}

Inserting (\ref{nfcn1}) in (\ref{nfcn2}), one gets

\begin{eqnarray}
\lefteqn{- 2 p_2 p_3 (a_{12} + a_{13}) -  a_{23} (p_2 a_{13}
 + p_3 a_{12}) >} \nonumber \\
& & p_1 p_2 p_3 - p_1 a_{23}^2
 - p_2 a_{13}^2 - p_3 a_{12}^2 - 2 a_{12} a_{13} a_{23}+ 2 p_1(p_2 p_3 -
 a_{23}^2)
\end{eqnarray}

which in view of (\ref{nfcbound}) cannot be satisfied. 

Thus, once again, the CP violating phase disappears at high temperature. As 
in the two-doublet case, here too the problem is that CP violation is
 achieved through the relative phase of the vevs of the doublets.

\subsection{$\; \not\!\!\!\!{\rm CP}$  with a singlet field}
\label{sin}

It should be clear from the previous examples that when the CP phase is 
related to the relative phases of doublet fields, high temperature effects 
will make it vanish. We therefore look for models in which CP violation 
is broken spontaneously by the vev of just one field, which may be easier 
to keep at high temperature.

The simplest such model is a minimal extension of the Standard Model with 
\begin{description}
\item[a)] a real singlet field S which transforms under CP as 
$S\rightarrow -S$.
\item[b)]an additional down quark, with both left and right components 
$D_L^a$ and $ D_R^a$ 
 singlets under $SU(2)$.  
 
\end{description}

The interaction Lagrangian for the down quarks, symmetric under CP, 
contains the terms

\begin{eqnarray}
{\cal L}_Y &=& (\bar{u} \bar{d} )^a_L h_{a} \Phi D_R + 
(\bar{u} \bar{d} )^a_L h_{ab} \Phi d^b_{R} \nonumber \\
& &  + M_D \bar{D_L} D_R + M_a( \bar{D_L} d^a_{R} + h.c.) \nonumber \\
& & + i f_D S (\bar{D_L} D_R - \bar{D_R} D_L)
+ i f_a S (\bar{D_L} d^a_{R} - \bar{d^a_{R}}D_L )
\label{sinyuk}
\end{eqnarray}

Clearly, when $S$ gets a vev (at a scale $\sigma$ much bigger than the 
weak scale $M_W$) CP is  spontaneously broken by the terms in the last line. 
A model of this
 kind was developed by Bento and Branco \cite{bb90}, in the version where 
the singlet is a complex field and gets a complex vev, and with an additional 
symmetry under which $S$ and $D_R$ are odd, all other fields even.
 
We will for simplicity keep $S$ real (and impose no further symmetries),
 noting that the analysis goes over the same lines as in \cite{bb90}, and
 referring the reader there for details. Suffice it to say that 
CP violation is achieved by complex phases appearing in the CKM
 matrix through the mixings of $d$ and $D$ quarks,  which are of the order
 $\sigma / M_D$. These phases remain in the limit $M_D, \sigma \to \infty$ 
when the heavy quarks 
decouple. This should not come as a surprise, since in the decoupling limit
 the theory  reduces to the minimal standard model, which in general has
 complex Yukawa couplings and a complex CKM matrix.
Also, flavor-violating currents are suppressed by powers of
 $M_W / \sigma$, disappearing in the decoupling limit. Thus the measure 
of the departure from the standard model is the dimensionless parameter
 $M_W/M_D$, and for the theory to be experimentally testable $M_D$ should 
not be much bigger than 1 TeV.

To leading order, the high-temperature behavior of the $\Phi-\sigma$ 
system is very simple. The most general potential can be written as

\begin{eqnarray}
V(\Phi, S) &= & - m_\Phi^2 \Phi^\dagger \Phi +
 \lambda_\Phi (\Phi^\dagger \Phi)^2 \nonumber \\
& & - {m_S^2 \over 2} S^2 + {\lambda_S \over 4}  S^4 - 
{\alpha \over 2}   \Phi^\dagger \Phi S^2
\label{sinpot}
\end{eqnarray}

and it has a minimum at

\begin{equation}
\langle \Phi \rangle = {1 \over \sqrt{2}}
\left(\begin{array}{c} 0\\v \end{array} \right)
\;\;\; ; \;\;\; \langle S \rangle = + \sigma
\end{equation}

At high T, the masses are replaced by

\begin{eqnarray}
m_\Phi^2(T) &=& - m_\Phi^2 + {T^2 \over 24}( 12 \lambda_\Phi - \alpha) 
\nonumber \\
{m_S^2(T) \over 2} &=&- {m_S^2 \over 2} + {T^2 \over 24} (3 \lambda_S - 2 
\alpha) 
\label{sinmass} 
\end{eqnarray}

We can  have $m_S^2< 0$ always  by 
requiring $2 \alpha > 3 \lambda_S$, and thus $\sigma \neq 0$ at any
 temperature.  The only further restriction is the usual $\lambda_\Phi >
 \alpha^2/\lambda_S$.

It seems then that in this model, one can have CP broken at any 
temperature. Remember however that up to now we have only considered 
 the leading order contributions to the effective potential in calculating 
the masses (\ref{sinmass}). A complete analysis should include the next-to 
leading order corrections, as we already mentioned in the Introduction. We
 can anticipate that for a singlet field these effects will not 
 change the picture much, but we leave a detailed analysis for a 
separate section.

\section{Spontaneous P violation and high T}
\label{p}

Spontaneous $P$ violation has been already discussed in the second paper 
of ref. \cite{ms79}, mostly in connection with strong CP violation. 
It was concluded there that in the minimal models of spontaneous P violation, 
left-right asymmetry may persist to high temperatures. The analysis 
however was carried out without  considering carefully the role of the gauge
 couplings, which is now known to be fundamental \cite{nos}, and which as we 
will show may invalidate that conclusion. 

Let us recall the salient features of the minimal left-right symmetric 
theories \cite{s79} based on a $SU(2)_L\times SU(2)_R \times U(1)_{B-L}$
 gauge symmetry. The fermions are in doublet representations

\begin{eqnarray}
\left(\begin{array}{c}u \\ d \end{array}\right)_L &\;\;\;\; ; \;\;\;&
 \left(\begin{array}{c}u \\ d \end{array}\right)_R \nonumber \\
\left(\begin{array}{c}\nu \\ e \end{array}\right)_L &\;\;\;\; ; \;\;\;&
 \left(\begin{array}{c}\nu \\ e \end{array}\right)_R 
\end{eqnarray}

The minimal Higgs sector of the theory consists of
\begin{itemize}
\item the  bi-doublets
 (one or more) $\Phi$ needed to provide Yukawa couplings and  fermion 
masses
\item two  multiplets $\Delta_L$ and $\Delta_R$ which may be either doublets
or triplets under $SU(2)_L$ and $SU(2)_R$, and which are in charge of
 breaking P spontaneously.
\end{itemize}

For the sake of completeness, we remind the reader of the essence of
spontaneous  P
 violation and we do it in a simplified toy example which has all the 
relevant features of the theory. More precisely, we take $\Delta_L$ and 
$\Delta_R$ as real scalar fields and assume a left-right symmetric potential

\begin{eqnarray}
V &=& -{m^2 \over 2}(\Delta_L^2 + \Delta_R^2) +
 {\lambda \over 4} (\Delta_L^4 + \Delta_R^4) + 
{\lambda' \over 2}\Delta_L^2\Delta_R^2 \nonumber \\
&=& -{m^2 \over 2}(\Delta_L^2 + \Delta_R^2) +
 {\lambda \over 4} (\Delta_L^2 + \Delta_R^2)^2 + 
{\lambda' - \lambda  \over 2}\Delta_L^2\Delta_R^2
\end{eqnarray}

A simple inspection of V is enough to convince oneself that for $m^2 >0$ and 
$ \lambda' - \lambda >0$, the global minimum of the theory is obtained for

\begin{equation}
\langle \Delta_L \rangle^2 = 0 \;\;\; ;\;\;\;\langle \Delta_R \rangle^2 = 
 {m^2 \over \lambda}
\end{equation}

or viceversa. Thus the left-right symmetry is broken spontaneously. Of
 course in realistic models, besides $\Delta$'s being non-trivial 
representations under the gauge group, we do need a field $\Phi$. 
One can then try to take one or more of the coupling constants between 
$\Phi$ and
 the $\Delta$'s negative, 
thus achieving a negative mass term  for the $\Delta$'s at all temperatures. 

Let us concentrate in the version of the theory which incorporates the
 see-saw mechanism with $\Delta_L$ and $\Delta_R$ being triplets \cite{ms81}. 
Since
 we wish to keep $\langle \Delta_R \rangle$ nonzero at high temperature,
 it is enough to look at the $\Delta_R -\Phi$ system and, as
 in \cite{ms79}, consider a simplified model in which the potential  
is written

\begin{eqnarray}
V = - m^2_\Delta \Delta_R^\dagger\Delta_R +
 \lambda_\Delta (\Delta_R^\dagger\Delta_R)^2 +
& &- m^2_\Phi Tr \Phi^\dagger\Phi +
 \lambda_\Phi (Tr \Phi^\dagger\Phi)^2 - 
2 \alpha Tr \Phi^\dagger\Phi \Delta_R^\dagger\Delta_R
\label{lrpot}
\end{eqnarray}

where $\Delta_R$ is a triplet under $SU(2)_R$, has $B-L$ number 2, and 
other couplings are taken to be small. The high temperature masses are
\footnote{We use the normalization $Tr \Phi^\dagger\Phi = \Phi_a \Phi_a/2$; $
 \Delta_R^\dagger\Delta_R = \Delta_R^a\Delta_R^a$, where $a$ 
 sums over six real fields.} 

\begin{mathletters}
\label{lrmass}
\begin{equation}
m_\Phi^2(T) = -m_\Phi^2 + T^2\left\{ {5\over 6}\lambda_\Phi - {1 \over 3}
 \alpha  + {3 \over 16} g^2 \right\}
\end{equation}
\begin{equation}
m_\Delta^2(T) = -m_\Delta^2 + T^2\left\{ {1\over 2}\lambda_\Delta - 
{2 \over 3} \alpha  + {3 \over 8} (g'^2 + 2 g^2)
 \right\}
\end{equation}
\end{mathletters}

where $g'^2$ is the $U(1)$ gauge coupling, $g^2$  the $SU(2)_R$ one. 
We have to keep $m_\Delta^2(T)$ negative at high T while preserving the
 boundedness condition $\lambda_\Phi\lambda_\Delta >\alpha^2$, thus we
 arrive at

\begin{eqnarray}
\lambda_\Phi &>& {\alpha^2\over \lambda_\Delta} \nonumber \\
&> &
{9 \over 4}\left[ {1\over 2}\lambda_\Delta + 
{3 \over 8} (g'^2 + 2 g^2)\right]
\end{eqnarray}

$\lambda_\Phi$ as a function of $\lambda_\Delta$ has a minimum at 
$\lambda_\Delta = (3/4) (g'^2 + 2 g^2)$, so we must have

\begin{equation}
\lambda_\Phi > {27 \over 16} (g'^2 + 2 g^2)
\label{pcond}
\end{equation}

If we now use $g'^2 = g^2/2$ and take $g^2 = 1/4$, we see that nonrestoration 
of P requires $ \lambda_\Phi > 1 $ in conflict with perturbation theory.  
Including other couplings does not help, since new conditions on the
 couplings coming from the mass matrices have to be imposed (since it is 
not illustrative, we omit here the numerical analysis required to prove this).

Although physically less attractive, one can in principle use doublets to
 break P spontaneously. This is actually the case studied in \cite{ms79}.
It is easily found that with doublets 
 the condition equivalent to (\ref{pcond}) is  down by a factor
 of half. Thus this case may be considered borderline.

Now, for the implementation of the see-saw mechanism in its minimal form,
 it turns out that a parity odd singlet field is needed \cite{cm85}. 
The singlet field $S$ will couple to the  $\Delta$ fields 
with a left-right symmetric term

\begin{equation}
  M S ( \Delta_L^\dagger \Delta_L - \Delta_R^\dagger \Delta_R)
\end{equation}

Without the lower
 bound imposed by the gauge couplings, the situation in this case 
 goes along the same lines as that of section \ref{sin}: the vev of the 
singlet can be kept nonzero at high temperatures with the aid of the 
bi-doublet field $\Phi$, or even of the $\Delta$'s.
Exactly as it worked with CP, now P may remain broken at high temperature,
 and the presence 
of more fields coupled to $S$ than in the CP case only makes it easier. 

\section{Strong CP Problem and High T}
\label{strongcp}

The strong CP problem arises in QCD when nonperturbative effects, resulting 
from the existence of instanton solutions, induce effective terms in the 
Lagrangian that violate CP. The resulting CP violating phase is

\begin{equation}
\bar\Theta = \Theta + {\rm arg}\, {\rm det}(M)
\end{equation}

where $\Theta$ is the coefficient of the
 $\epsilon_{\alpha\beta\mu\nu}F^{\alpha\beta}_a F^{\mu\nu}_a$ term,
 and $M$ is the quark's mass matrix.
$\bar\Theta$ is  constrained experimentally to be zero to a very high precision
 ($\bar\Theta < 10^{-9}$), giving rise to a ``naturalness'' problem \cite{k87}.

\subsection{The invisible axion solution}

The most popular solution to  the  strong CP problem is the 
Peccei-Quinn mechanism \cite{pq77},
 in which the phase $\bar\Theta$ is identified with   
the pseudo-Goldstone boson resulting 
from the spontaneous breakdown of a global symmetry $U(1)_{PQ}$. 
Observational constraints require this breakdown to occur at a scale $M_{PQ}$
much bigger than the electroweak scale, making the axion ``invisible'' 
\cite{k79,svz80}. Besides the axion field $a$,
 the breaking of $U(1)_{PQ}$ produces a network of global strings \cite{s82}.
 As we go around each minimal  string, the phase $\bar \Theta = a/M_{PQ}$ 
winds by $2\pi$.
Instanton effects appear later, when the temperature has reached the QCD
 scale $\Lambda_{QCD}$.  Their effects in the Higgs sector can be mimicked by
an effective term

\begin{equation}
\Delta V = \Lambda_{QCD}^4(1-cosN\bar{\Theta})
\end{equation}
where $N$ is the number of quark flavors.
 It becomes energetically favorable
 for $\bar{\Theta}$ to choose one out of the discrete
set of values $2\pi k/N$ ($k=1,2,..N$). But since we must have 
$\Delta \bar{\Theta} =2\pi$ around a string, this results 
in the formation of $N$
 domain walls attached to each string \cite{ve82}. For $N >1$, these 
domain walls are stable and therefore in conflict with standard cosmology. 

Clearly, without the global strings no walls will be formed: above 
$T\simeq \Lambda_{QCD}$, $\bar\Theta$ would be aligned having some typical
value $\bar\Theta_0$ which after the QCD phase transition would relax to 
the nearest minimum. 
We wish then to study in detail the high temperature behavior of 
the invisible axion mechanism,  well above the scale $M_{PQ}$. 

For concreteness we concentrate on the minimal extension of the original
 Peccei-Quinn model \cite{svz80}.
The potential for the PQ model with the
 doublets $\phi_i $ (i=1,2) both having $Y=1$ and a $SU(2) \times U(1)$ 
singlet  $S$ may be written as

\begin{eqnarray} 
V_{PQ} &=& \sum_i{\left[-{m_i^2 \over 2}  \phi_i^\dagger \phi_i + 
{\lambda_i \over 4} (\phi_i^\dagger \phi_i)^2\right]} 
-  {\alpha \over 2}(\phi_1^\dagger \phi_1)(\phi_2^\dagger \phi_2)-
 {\beta \over 2}(\phi_1^\dagger \phi_2)(\phi_2^\dagger \phi_1) \nonumber \\
&-& {m_s^2 \over 2} S^* S + {\lambda_s \over 4} (S^* S)^2 
- \sum_i{({\gamma_i \over 2} \phi_i^\dagger \phi_i}) S^* S
 - M (\phi_1^\dagger \phi_2 S + \phi_2^\dagger \phi_1 S^*) 
\label{pqpot}
 \end{eqnarray}

Besides the $SU(2)_L \times U(1)_Y$ local gauge symmetry, $V_{PQ}$ has a
 chiral $U(1)_{PQ}$ symmetry ($\phi_1$ couples to say down quarks, and
 $\phi_2$
to up quarks)

\begin{equation}
\phi_1 \rightarrow e^{i \alpha }\phi_1 \;\; ; \; \; \; \phi_2 \rightarrow 
e^{- i \alpha }\phi_2 \;\; ; \; \; \; S \rightarrow e^{2 i \alpha }S
\label{pqsim}
\end{equation}

For $\beta >0$, the minimum is found at

\begin{equation}
\langle \Phi_i \rangle =\left(\begin{array}{c}
0 \\v_i \end{array}\right)  \;\; \; ; \;\;\;
 \langle S \rangle = v_S
\label{pqvev}
\end{equation}

To have $U(1)_{PQ}$ broken at any temperature, it is enough to keep the vev
 of the singlet nonzero for all T. From our analysis of the previous
 section for a potential with three doublets, one can already expect that
 keeping the vev of only one field nonzero   will not be difficult. 
In this  model then the conditions on the potential parameters cannot be 
an obstacle for nonrestoration, but we present them here for the sake of
 completeness. Taking $v_S \gg v_i$, the conditions over the couplings are,
 to leading order

\begin{mathletters}
\label{pqcond}
\begin{equation}
\lambda_i >0 \;\;\;, \;\; \lambda_S>0 \;\;\; ;\;\; 
\lambda_i\lambda_S>\gamma_i^2 \;\;\; ;\;\;
 \lambda_1\lambda_2 > (\alpha +\beta)^2 
\label{pqcond1}
\end{equation}
\begin{eqnarray}
& &Mv_s^3\left[{v_1^3 \over v_2}(\lambda_1\lambda_S-\gamma_1^2) +
{v_2^3 \over v_1}(\lambda_2\lambda_S-\gamma_2^2) 
- 2 v_1v_2 (\lambda_S(\alpha+\beta) + \gamma_1 \gamma_2\right] \nonumber \\
&+ & v_S^2 v_1^2 v_2^2 \left[\lambda_1\lambda_2\lambda_S - 
\lambda_1\gamma_2^2 - \lambda_2\gamma_1^2  -\lambda_S (\alpha+\beta)^2
- 2 \gamma_1\gamma_2(\alpha+\beta)\right] >0
\label{pqcond2}
\end{eqnarray}
\end{mathletters}

It is easily proven that (\ref{pqcond1}) imply that the first line of eq. 
(\ref{pqcond2}) is positive. A sufficient condition for boundedness will
 then require  (\ref{pqcond1}) and the second line of (\ref{pqcond2})
 to be positive, the same conditions that were required in the three-doublet
 model of section \ref{nfc} (eq.(\ref{nfcbound})).

The mass term of the singlet at high temperature will be

\begin{equation}
m_S^2(T) = - m_S^2 + {T^2 \over 3} (\lambda_S -\gamma_1 -\gamma_2)
\label{pqmass}
\end{equation}

so that imposing $\gamma_1 +\gamma_2 > \lambda_S$, we get the $U(1)_{PQ}$ 
symmetry broken at all temperatures. We already know that at high T one
 cannot have all three vevs nonzero, and notice that because of the 
linear terms in (\ref{pqpot}), having $v_S \neq 0$ forces $v_1, v_2 $ 
to vanish. 

Up to this order then, it seems quite natural to keep the vev of $S$ 
nonzero at high T; again we leave the next-to-leading order considerations
 for the next session. The learned reader will notice that the same holds 
true for  Kim's  version \cite{k79} of the invisible axion idea. 

\subsection{Spontaneous P or CP violation}

Another well-known solution to the strong CP problem is based on the idea 
of spontaneous CP or P violation \cite{h78}. Here, the symmetries can be 
used to set
 $\bar\Theta_{\rm tree}=0$ and the effective $\bar\Theta$ is then
 finite and calculable in perturbation theory, and in many models 
   small enough. The high T behavior of these theories is completely
 analogous to the one discussed in section \ref{weakcp} and \ref{p}, and
 thus we can conclude that the solution of the domain wall problem favors
 models with singlets. However, before {\em the} model is found we find it
 fruitless to study this question in detail. 
 
\section{Next-to-leading Order Contributions}
\label{ntl}

In a series of recent papers, Bimonte and Lozano  \cite{bl95a,bl95b}
have addressed the issue of next-to-leading order contributions to the 
effective potential. As was already pointed  out in \cite{dj74}, 
in a theory with a $\lambda \phi^4$ 
potential, the next-to-leading order contributions to the mass$^2$ are of order

\begin{equation}
m^2(T) \propto \lambda^{3/2} T^2
\end{equation}

while higher loop corrections do not contribute significantly. The point 
is that in a theory with two fields where one of the self-coupling constants 
is required to be larger than the other (as we did to avoid symmetry
 restoration), the larger constant will enter in corrections to the other 
field's mass. Thus one has to make sure that the results to leading order 
are maintained when including such terms. 

In fact, in the case of gauge symmetries, it was concluded \cite{bl95b} that
 the inclusion of these effects can alter significantly the phase diagram of
 the theory. This is mainly due to the fact that in the gauge case the
 coupling constants cannot be as small as one wishes, but are bounded
 from below by the value of the gauge coupling. In the case of singlets 
 \cite{bl95a}, although the effects are not so dramatic, they do alter 
the parameter space for symmetry nonrestoration. Since in this investigation 
the models that allow for  nonrestoration at high T were based on singlet
 fields, we will only consider here the next-to-leading order corrections 
in the case of global symmetry. 

We begin by reviewing briefly   the contributions of next-to-leading 
corrections  in  the effective potential of a 
$O(N_1)\times O(N_2)$-symmetric model, although we refer the reader to 
\cite{bl95a} for details. Take two real fields $\phi_1, \phi_2$, transforming 
as vectors under $O(N_1),O(N_2)$ respectively, and write the potential

\begin{equation}
V(\phi_1,\phi_2) = \sum_i\left( -{m_i^2 \over 2} |\phi_i|^2 + 
{\lambda_i\over 4} |\phi_i|^4 \right) - {\alpha \over 2} |\phi_1|^2|\phi_2|^2
\end{equation}

The  temperature contributions to the effective masses are calculated 
to leading order to be

\begin{equation}
\Delta m_1^2(T) = T^2 \nu_1^2  = T^2 \left[\lambda_i
 \left({2 + N_1 \over 12} \right) - {N_2 \over 12} \alpha \right]
\end{equation}

(and a similar expression for $\Delta m_2$) while to next-to-leading,
 $\Delta m_i \equiv  T x_i$ is found by solving the 
coupled pair of equations

\begin{eqnarray}
x_1^2 &=& \nu_1^2 - \left({2+N_1 \over 4 \pi}\right)\lambda_1 x_1 + 
{N_2 \over 4 \pi} \alpha x_2 \nonumber \\ 
 x_2^2 &=& \nu_2^2 - \left({2+N_2 \over 4 \pi}\right)\lambda_2 x_2 + 
{N_1 \over 4 \pi} \alpha x_1 
\end{eqnarray}

Symmetry is restored when such solutions are real and positive. 
The conditions under which those solutions do not exist, and therefore 
the $O(N_2)$ symmetry is {\em not} restored 
can be found to be

\begin{mathletters}
\label{ntlcond}
\begin{equation}
\alpha \left({N_1 \over  2 + N_2}\right) [ 1 - f(\lambda_1,\alpha)] > \lambda_2
\label{ntlcond1}
\end{equation}
\begin{equation}
\lambda_1 \lambda_2 > \alpha^2
\label{ntlcon2}
\end{equation}
\end{mathletters}

where

\begin{equation}
f(\lambda_1,\alpha) = {3 (2+N_1) \over 8 \pi^2} \left(
\sqrt{\lambda_1^2 + \left({16 \pi^2 \over 3 (2+N_1)} \right)
\left(\lambda_1 - {N_2 \over 2+N_1}\alpha \right)}
- \lambda_1  \right)
\end{equation}

is a function that can take values from $0 $ to $1$.
The leading order conditions are (\ref{ntlcond}) with $f=0$. One can see 
then why the parameter space is reduced: it gets more  difficult to fulfill
(\ref{ntlcond1}). The behavior with the number of fields also becomes 
nontrivial, since  $(1 -f)$ is a decreasing function of
 $N_1$, and the two factors of $\alpha$ in (\ref{ntlcond1}) compete (up
 to leading order, it is always preferable to keep nonzero the vev of the 
field in the smallest representation).

The $O(N_1)\times O(N_2)$ toy model can mimick models with more complicated 
symmetries involving two fields with $N_1$ and $N_2$ real components, in the
approximation where their interaction is just of the type $\alpha 
|\phi_1|^2|\phi_2|^2$. In particular, no approximation needs to be done in
 the doublet+singlet case.

\begin{figure}
\centerline{\psfig{figure=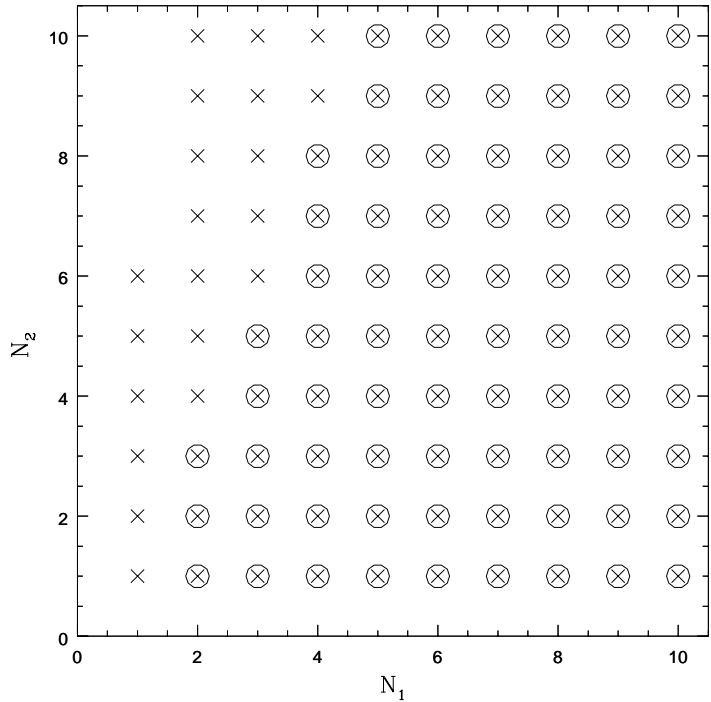,height=10cm}}
\caption{Symmetry nonrestoration in a model with $O(N_1)\times O(N_2)$
 symmetry. Points indicate the values of $N_1$, $N_2$ for which the vev 
of the $O(N_2)$ vector can be kept nonzero at high temperature, for 
fixed values of the potential's parameters: circles correspond
 to $\lambda_1=0.1, \, \alpha = 0.03, \, \lambda_2 = 0.01$, crosses 
to $\lambda_1=0.1, \, \alpha = 0.01, \, \lambda_2 = 0.001$}
\label{n1n2}
\end{figure}

In Figure \ref{n1n2}  we show how symmetry nonrestoration
 depends in the number of fields when the next-to-leading order effects are
 included, i.e., we find the values of $N_1$ and $N_2$ for which the 
conditions (\ref{ntlcond}) are satisfied when the parameters of the potential
 are fixed. 
The plot shows the situation for 
two sets of ratios of the couplings: $\lambda_1:\alpha:\lambda_2 = 1:1/3:1/9$
 and $ 1:1/10:1/100$.
Notice that $N_2 < N_1$ is still preferred. As the ratio $N_2/ N_1$
 increases,
it becomes necessary for nonrestoration to take smaller ratio
 $\lambda_2/\lambda_1$. 

 The cases of $N_1=4, N_2 = 1$
 (a complex doublet  plus a real singlet, as required for CP violation in 
section \ref{sin}), that of $N_1=8, N_2=2$ (two doublets and one complex
 singlet, as in the invisible axion model of section \ref{strongcp}) 
and that of $N_1=8, N_2=1$ (two doublets and a singlet, as in the
 parity-violating model of section \ref{p}) lie 
in the non-restoration region. 

\begin{figure}[h]
\centerline{\psfig{figure=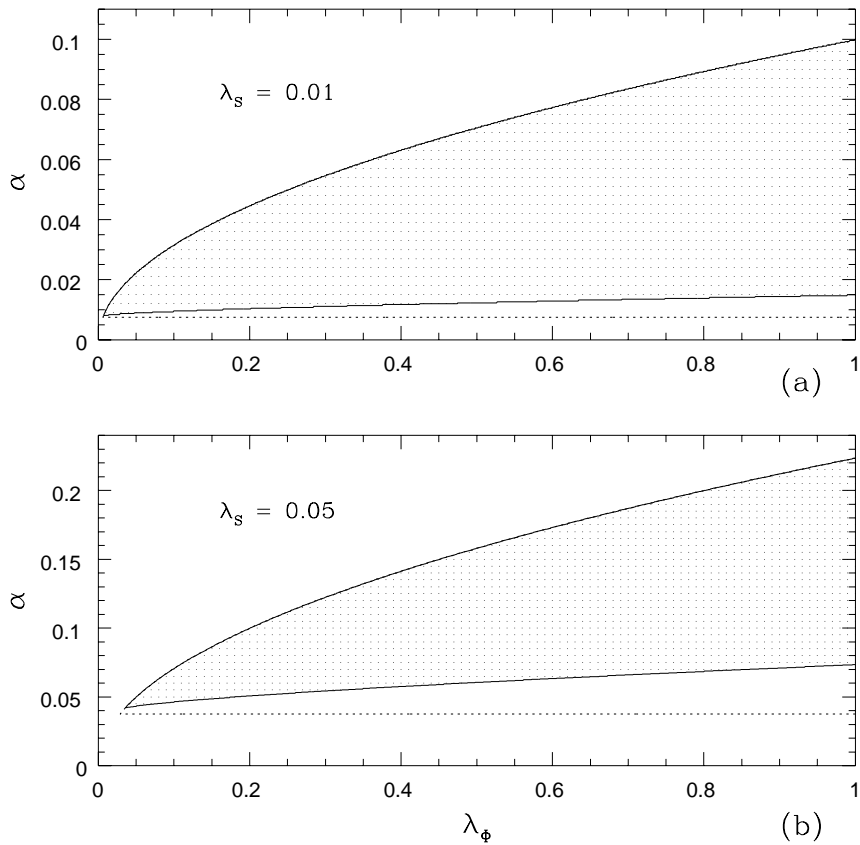,height=12cm}}
\caption{ The region of symmetry nonrestoration for the model of 
   $\not\!\!\!\!{\rm CP}$ with a real, CP odd singlet, for two values 
of the singlet's 
self coupling constant $\lambda_s$ as indicated.
 When only leading order effects
 are taken into account, the region extends up to the dotted line}
\label{bb}
\end{figure}

The relevant question is how big is the region in parameter space where 
nonrestoration occurs. In Figure \ref{bb} we show that region for the case of 
the CP violation with a real singlet, in $\lambda_\Phi, \alpha$ space, when 
$\lambda_S$ is kept at a fixed value. Varying $\lambda_S$ basically
 `rescales' the whole picture in the $\alpha$ axis. The corresponding 
region with only leading-order effects is also shown. Although the 
parameter space is reduced by higher order corrections, the difference 
with the leading order case is not dramatic.

For the Peccei-Quinn model, the next-to-leading order calculations
are only approximated by an  $O(8) \times O(2)$ model, in the limit
 where in (\ref{pqpot}), $\lambda_1 = \lambda_2 = 2 \alpha \equiv 
\lambda_\Phi$, 
$ \beta =0$, and $\gamma_1 = \gamma_2 \equiv \gamma$. 

\begin{figure}[h]
\centerline{\psfig{figure=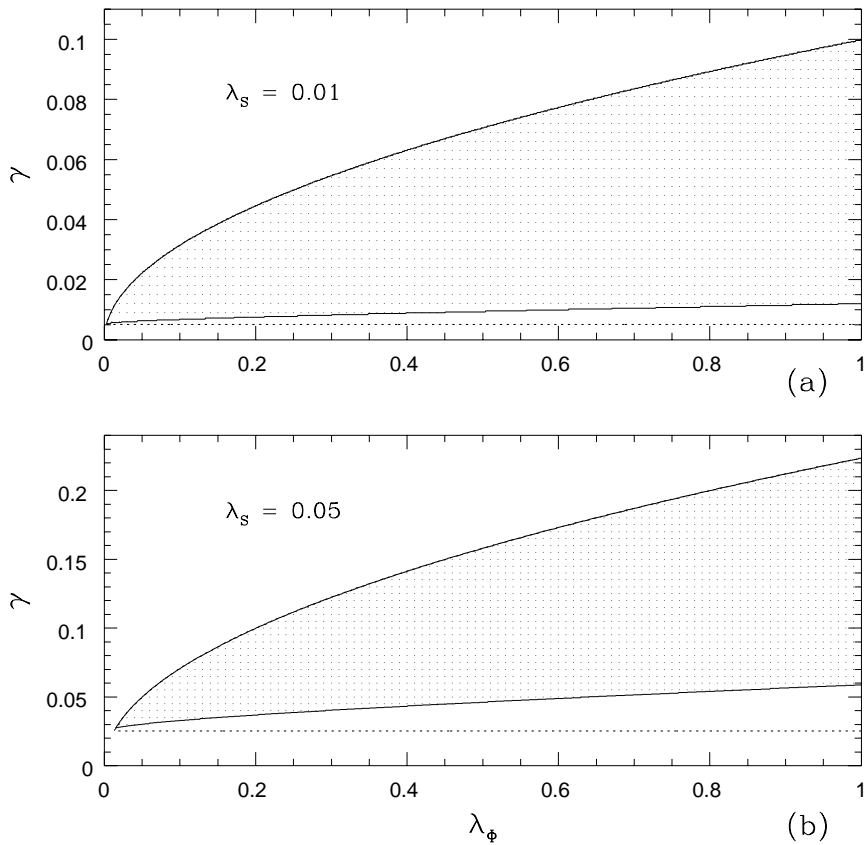,height=12cm}}
\caption{ The region of symmetry nonrestoration for a $O(8) \times O(2)$
model, an approximation of the  Peccei-Quinn model}
\label{pq}
\end{figure}

Under such approximation, the region where nonrestoration is allowed 
is presented in Figure \ref{pq}, for the same range of parameters as in 
Figure 2.
It is evident comparing both figures that nonrestoration does not depend 
only on the ratio $N_2/N_1$.

As for the model of $P$ violation with a singlet of section \ref{p}, it 
can be imitated by a $O(8)\times O(1)$ model if the quartic coupling with 
 the two doublet  fields is taken negative. One can also choose
 the couplings with the bi-doublet negative, and then consider an
 approximated model with some of the self and mixed  couplings small.  
The nonrestoration  region is clearly bigger than in the  weak or 
strong $CP$ cases. 

\section{Outlook and Conclusions}

In this paper we have studied the phenomenon of symmetry nonrestoration 
at high temperature, focusing on some minimal models of spontaneous T and P
 violation. We were motivated by the fundamental role that these symmetries
 play in nature and by the possibility of using them in solving the strong
 CP problem. We find that symmetry nonrestoration seems to require singlet 
fields and that it seems to work in accord with perturbation theory. This
 provides the hope for solving the domain wall problem and having 
baryogenesis operate at very high temperature as we now discuss briefly. 

\vspace{0.5cm}
{\bf Domain Wall Problem}:\hspace{0.5cm} Avoiding the phase transition
is not enough to  solve  the domain wall
problem, since thermal fluctuations are in principle able to produce 
topological defects  at any time. 
As was shown in \cite{ds95}, thermal production
 of domain walls and strings  can be naturally suppressed. We briefly 
sketch how this suppression occurs for the two models admitting 
nonrestoration presented here, and refer to \cite{ds95} for details.

Consider the nucleation of a large spherically symmetric domain wall or
 a closed loop of string. The production  rate per unit time per unit volume 
at a temperature $T$ will be given by \cite{l81}

\begin{equation}
\Gamma = T^4 \left({S_3\over 2\pi T }\right)^{3/2} e^{-S_3/T}
\label{rate}
\end{equation}

where $S_3$ is the energy of the closed defect. The suppression factor 
$e^{-S_3/T}$ is readily calculated in the limit where the defect's radius 
is much bigger than its width.  For the domain walls produced in the model 
of CP violation with a singlet, we get

\begin{equation}
{S_3 \over T} \gg {16 \pi \over 3 \sqrt{6}} {\sqrt{2\alpha - 3\lambda_S}
\over \lambda_S}
\label{sup1}
\end{equation}

 Analogously, for the Peccei-Quinn model the thermal production of 
large loops of strings is suppressed by \footnote{We note that
 the normalization of the kinetic term we use here differs from that 
of  \cite{ds95}.}

\begin{equation}
{S_3 \over T} \gg 
4 \pi^2 {\sqrt{\gamma_1 + \gamma_2 - \lambda_S} \over \lambda_S}
\label{sup2}
\end{equation}

We see that in both cases, it suffices to take the singlet's
 self-coupling $\lambda_S$ small to avoid significant thermal production
 of defects.

 The considered models with singlets involve a high scale $M_H$ much
 bigger than the
 weak scale $M_W$, and it is noteworthy that the smallness of $\lambda_S$ 
is intimately related to this hierarchy. Strictly speaking one could just
  fine tune the combination of $m_S^2$ and $\lambda_S v_S^2$ to be small,
 but this is not stable under radiative corrections. It is maybe more natural
 to take all the mass parameters of the model $m_\Phi$ and $m_S$ to
be small, i.e. of order $M_W$, and  the singlet's self and mixed couplings 
of order $(M_W/v_S)^2$. In such case it is obvious that both (\ref{sup1})
 and (\ref{sup2}) become enormous, suppressing completely the production
 of defects. Of course, the nature of the fine-tuning is finally a matter
 of taste. However, the second possibility has the clear prediction of
 keeping both Higgs doublets light in the invisible axion model, as is
 commonly assumed and experimentally verifiable.

Of course, all the above still does not guarantee the absence of domain 
walls. One needs to assume initial conditions in which the singlet field 
has a uniform value  over a region of roughly the comoving size of the
present
horizon. This is equivalent to assume that the so-called horizon problem 
has been solved, for example by means of a period of primordial inflation.

\vspace{0.5cm}
{\bf Baryogenesis}

The issue of baryogenesis in the context of  broken symmetries at high T
 has been discused in \cite{ms80} with emphasis on the theories where 
the $SU(2)\times U(1)$ gauge symmetry of the standard model never gets
 restored. This implies massive fermions at high T, but it can still be
 shown that baryogenesis may take place along the usual lines of the
 out-of-equilibrium decays of superheavy lepto-quark gauge and Higgs bosons.

Now, in the examples we have discussed both with P and CP violation
 at high T, and including the Peccei-Quinn mechanism, the $SU(2)\times U(1)$ 
symmetry gets restored as in the more conventional scenarios. Thus fermions 
become massless and 
the creation of baryon asymmetry proceeds as usual. Of
 course, this implies embedding of the models discussed into GUTs, a task
 beyond the scope of our paper.

\end{document}